\documentclass[runningheads]{llncs}
\usepackage{url}

\usepackage{cite}
\usepackage{graphicx}
\usepackage{subfigure}
\usepackage{amsmath}
\usepackage[noend]{algpseudocode}
\usepackage{algorithmicx,algorithm}
\usepackage{float}
\usepackage{booktabs}
\usepackage{multirow}
\usepackage{hyperref}
\usepackage{xcolor}

\usepackage{xcolor}
\newcommand{\hys}[1]{\textcolor{black}{#1}}
\begin{document}
\title{Infrared Image Super-Resolution via Heterogeneous Convolutional WGAN}
%
%
\author{Yongsong Huang\inst{1} \orcidID{0000-0003-3114-9206} \and Zetao Jiang \inst{1} \orcidID{0000-0002-0914-2131}
Qingzhong Wang \inst{2},\inst{3} \orcidID{0000-0003-1562-8098} \and Qi Jiang\inst{1} \and Guoming Pang\inst{4}}
\authorrunning{Y. Huang et al.}
%
\institute{Guilin University of Electronic Technology University, Guilin, China\\
\email{zetaojiang@guet.edu.cn\thanks{ Corresponding author: Zetao Jiang}},\email{hyongsong.work@gmail.com}, 
\and
City University of Hong Kong, Hong Kong SAR of China\\
\email{qingzwang2-c@my.cityu.edu.hk} 
\and 
Baidu Research, Beijing, China.
\and
ZTE Corporation, Shenzhen, China\\
\email{pang.guoming@zte.com.cn}
}
\maketitle              
\begin{abstract}
Image super-resolution is important in many fields, such as surveillance and remote sensing. However, infrared (IR) images normally have low resolution since the optical equipment is relatively expensive. Recently, deep learning methods have dominated image super-resolution and achieved remarkable performance on visible images; however, IR images have received less attention. IR images have fewer patterns, and hence, it is difficult for deep neural networks (DNNs) to learn diverse features from IR images. In this paper, we present a framework that employs heterogeneous convolution and adversarial training, namely, heterogeneous kernel-based super-resolution Wasserstein GAN (HetSRWGAN), for IR image super-resolution. 
The HetSRWGAN algorithm is a lightweight GAN architecture that applies a plug-and-play heterogeneous kernel-based residual block. Moreover, a novel loss function that employs image gradients is adopted, which can be applied to an arbitrary model. The proposed HetSRWGAN achieves consistently better performance in both qualitative and quantitative evaluations. According to the experimental results, the whole training process is more stable.
\end{abstract}
\section{Introduction}
Image super-resolution (SR) reconstruction is a very active topic in computer vision as it offers the promise of overcoming some of the limitations of low-cost imaging sensors. Infrared (IR) image super-resolution plays an important role in the military and medical fields and many other areas of vision research. A major problem with IR thermal imaging is that IR images are normally low resolution since the size and precision of IR sensors can be limited.
Image super-resolution is a promising and low-cost way to improve the resolution and quality of IR images. Generally, image super-resolution methods based on deep learning can be classified into two categories, namely, models based on generative adversarial networks (GANs)\cite{radford2015unsupervised,ledig2017photo} and models based on deep neural networks (DNNs)\cite{zhang2019deep,hui2019lightweight,haris2018deep,dong2015image,dong2016accelerating,shi2016real,nah2017deep,zhang2018learning}, 
both of which have achieved satisfying results on visible images. 
These methods can achieve a good peak signal-to-noise ratio (PSNR). 
However, they do not consider the visual characteristics of the human eye. The human eye is more sensitive to contrast differences with a lower spatial frequency. The sensitivity of the human eye to differences in brightness contrast is higher than its sensitivity to color, and the perception of a region by the human eye is affected by the surrounding areas. Situations in which the results of the evaluation are inconsistent with the subjective feeling of a viewer therefore often occur. We recommend using the structural similarity index (SSIM).
The learning-based SISR algorithm learns a mapping between low-resolution (LR) and high-resolution (HR) image patches. The prior knowledge used is either explicit or implicit, depending upon the learning strategy. The super-resolution convolutional neural network (SRCNN)\cite{dong_image_2016} algorithm introduced deep learning methods to SISR. A faster model, the faster super-resolution convolutional neural network (FSRCNN)\cite{dong2016accelerating}, improved upon the SRCNN model and has also been applied to SISR. The efficient subpixel convolutional neural network (ESPCN) algorithm\cite{shi2016real} and information multi-distillation network (IMDN)\cite{hui2019lightweight} were also proposed to further improve the computational efficiency. A significant advance in the generation of visually pleasing results is the super-resolution generative adversarial network (SRGAN)\cite{ledig2017photo}. A large number of SR methods have been presented, most of which are designed for natural images. Fewer methods have been designed for infrared images. GANs provide a powerful framework for generating plausible-looking natural images. However, they have problems with instability\cite{huang2021infrared,wang2018esrgan}. Wasserstein generative adversarial networks (WGAN)\cite{arjovsky_wasserstein_2017} was proposed as a solution to this problem. Given the issues that there are few infrared image features and that super-resolution reconstruction is difficult, the building units of the neural network and the loss functions that provide better constraints each play an important role in improving the performance of the GAN.

In this paper, we propose a novel approach for infrared image super-resolution. We revisited the key components of SRGAN and improved the model in two ways. First, we improved the network structure by introducing the heterogeneous kernel-based residual block, which has fewer parameters than previous algorithms, and it is easier to train. HetConv enables multiscale extraction of image features by combining convolutional kernels of different sizes. Second, we developed an improved loss function: the gradient cosine similarity loss function. The traditional loss function does not consider the characteristics of infrared images, and the gradient cosine similarity loss function takes the image gradient as an important feature for better-supervised training. The experimental datasets are publicly available\cite{huang_huang_2019}, and the experimental effects can be validated.

The remainder of this paper is organized as follows. The related works are presented in Section \ref{se.2}. We describe the HetSRWGAN architecture and the gradient cosine similarity loss function in Section \ref{se.3}. A quantitative evaluation of new datasets, as well as visual illustrations, is provided in Section \ref{se.4}. The paper concludes with a conclusion in Section \ref{se.5}.

\section{Related Works}\label{se.2}

\subsection{Generative Adversarial Networks}

Generative adversarial networks\cite{goodfellow_generative_2014} were proposed by Goodfellow, based on game theory. In a pioneering work, C. Ledig et al\cite{ledig2017photo} used SRGAN to learn the mapping from LR to HR images in an end-to-end manner, achieving performance superior to that of previous work. A low-resolution image $I^{L R}$ is input to a generator network to generate the reconstructed image $I^{SR}$, while a discriminator network takes the high-resolution images $I^{H R}$ and $I^{SR}$ as input to determine which is the real image and which is the reconstructed image. 
\subsection{HetConv: Heterogeneous Kernel-Based Convolutions}

\begin{figure}[t]
\centering
\includegraphics[width=\textwidth]{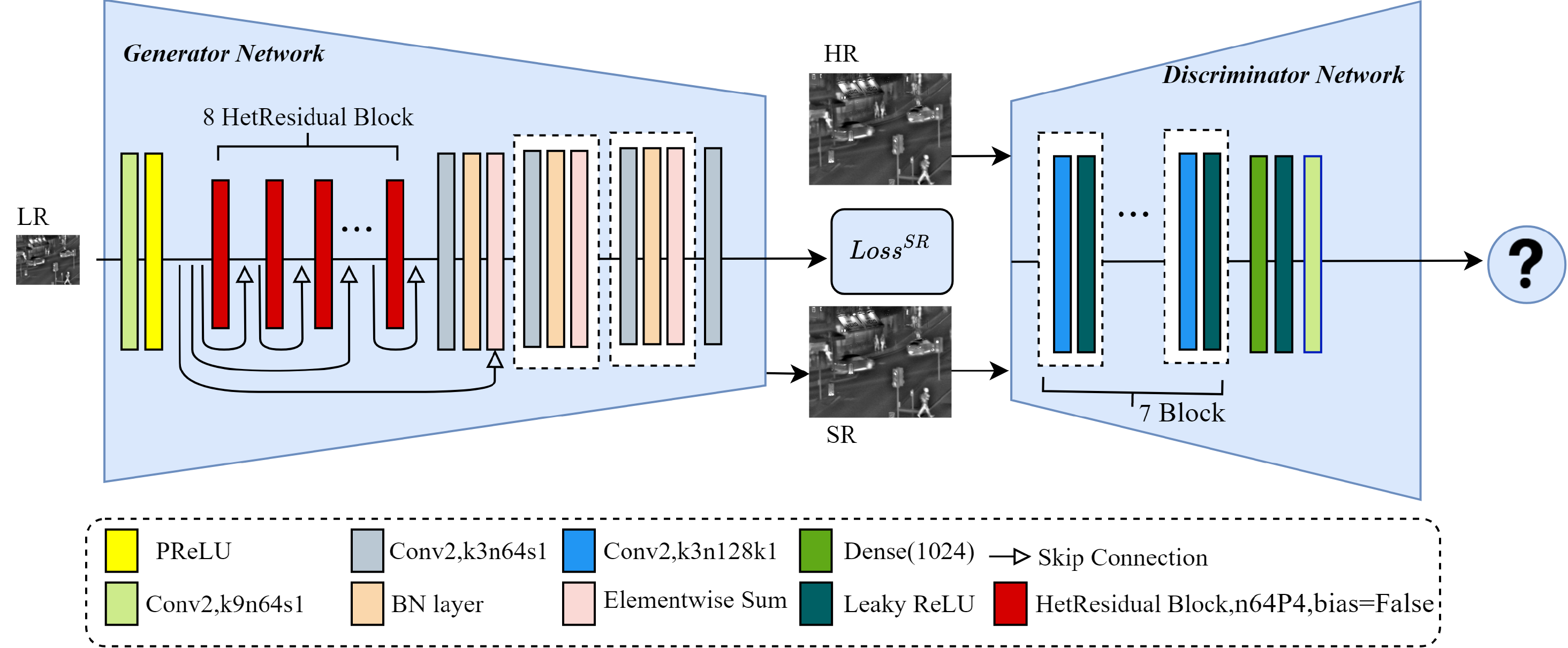}
\caption{Architecture of heterogeneous kernel-based super-resolution Wasserstein GAN with the corresponding kernel size ($k$), number of feature maps ($n$), stride ($s$) for each convolutional layer, padding ($p$) and number of the normal convolution kernel ($P$) (Best viewed in color).}
\label{fig.1}
\end{figure}

The heterogeneous kernel-based convolutions algorithm was proposed by Pravendra Singh\cite{singh_hetconv:_2019-1}. Pravendra Singh et al presented a novel deep learning architecture in which the convolution operation uses heterogeneous kernels. Compared to standard convolution operations, the proposed HetConv reduces the number of calculations (FLOPs) and parameters while still maintaining the presentation efficiency. HetConv is especially different from the depthwise convolutional filter used to perform depthwise convolution (DWC)\cite{chollet_xception:_2017-1}, the pointwise convolutional filter used to perform pointwise convolution (PWC)\cite{szegedy_going_2015} and the groupwise convolutional filter used to perform groupwise convolution (GWC)\cite{krizhevsky_imagenet_2012}. In HetConv, a variable $P$ is used to control how much of the normal convolution kernel is retained in the operation. In addition, the total reduction is $R$ for $K \times K$ kernels. The number of calculations of HetConv is compared with that of the normal convolution, as shown in Eq \ref{eq.1}.
\begin{equation}
R_{{HetConv}}=\frac{1}{P}+\frac{(1-1 / P)}{K^{2}}
\label{eq.1}
\end{equation}
According to the characteristics of the heterogeneous kernel-based convolutions, we used a skip connection when designing the generator network structure. The HetSRWGAN structure is shown in Figure \ref{fig.1}.

\section{HetSRWGAN}\label{se.3}

\subsection{HetSRWGAN Architecture}
Our main goal was to improve the overall visual performance of SR. In this section, we describe our improved network architecture. The main difference between the GAN and WGAN\cite{arjovsky_wasserstein_2017} is that the sigmoid function and batch normalization (BN)\cite{ioffe_batch_2015} layer of the discriminator network are removed. The entire neural network is stabilized by gradient punishment\cite{arjovsky_wasserstein_2017}. It has been shown that removing the BN layer improves performance and reduces complexity\cite{wang2018esrgan,nah2017deep}. Further, the removal of the BN layer contributes to improving the robustness of the network and reduces the computational complexity and memory consumption. We replaced the original basic block with a heterogeneous kernel-based residual block (HetResidual block), which includes HetConv, as depicted in Section \ref{se.3.2}. The HetResidual block is the basic network building unit. This block requires fewer parameters than the original basic block, improves network performance, and reduces computational complexity. More parameters may lead to a higher probability of mode collapse\cite{wang2018esrgan,huang2021infrared}, so reducing the total number of parameters is beneficial.
 For the discriminator network, we deepened the network structure and experimentally demonstrated that this modification improves image quality. The detailed experimental results are given in Section \ref{se.4}. According to the characteristics of the heterogeneous kernel-based convolutions, we used a skip connection when designing the generator network structure. 

\subsection{Heterogeneous Kernel-Based Residual Block}\label{se.3.2}

\begin{figure}[t]
\centering
\includegraphics[width=7 cm]{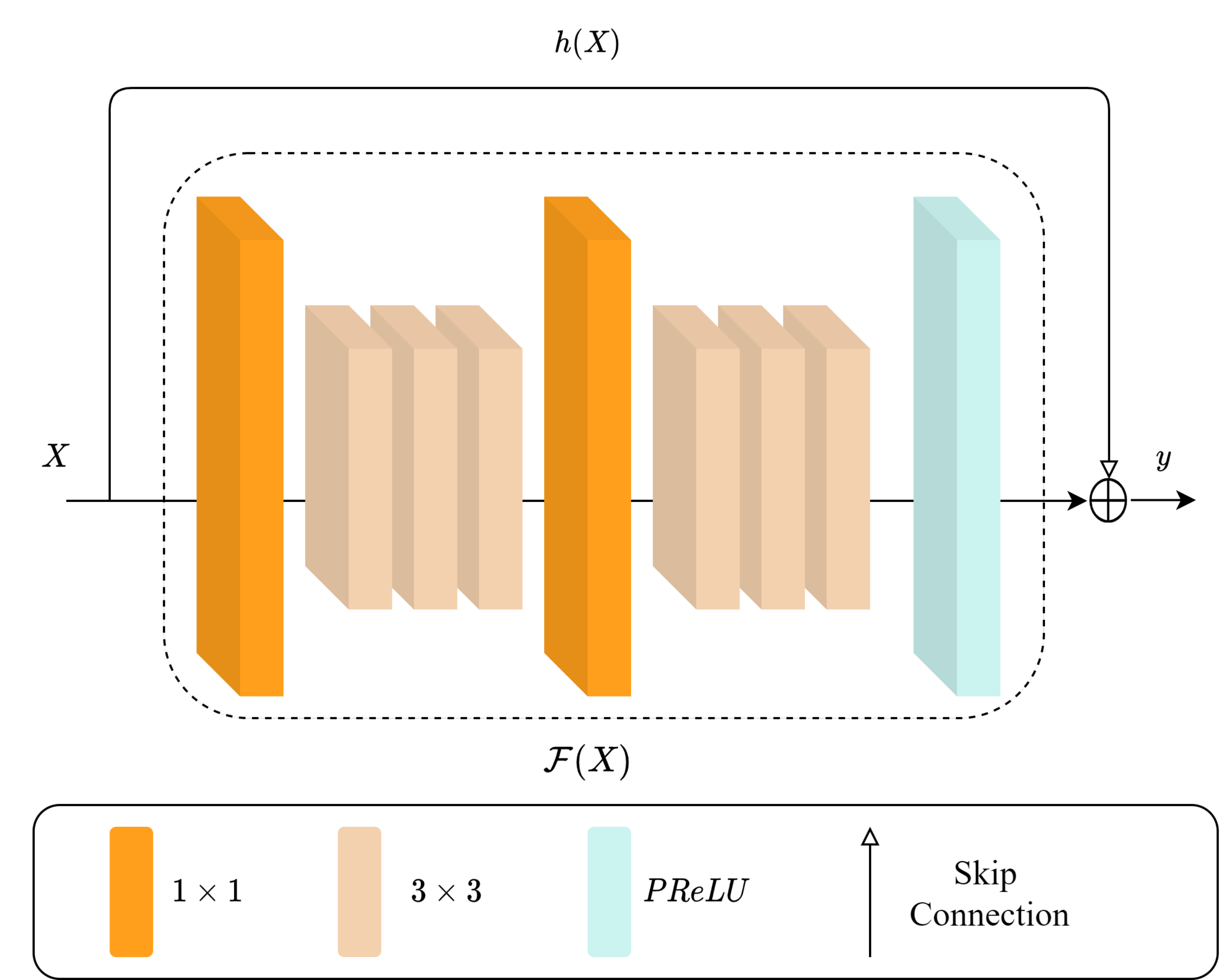}
\caption{Architecture of Heterogeneous Kernel-Based Residual Block.}
\label{fig.2}
\end{figure}

Kaiming He et al\cite{he_identity_2016} first proposed the residual block structure and solved some of the problems caused by deep neural networks by introducing a skip connection and combination. The heterogeneous kernel-based residual block is shown in detail in Figure \ref{fig.2}. The relevant formula is analyzed as follows:

\begin{equation}\mathbf{y}_{i}=h\left(\mathbf{x}_{i}\right)+\mathcal{F}\left(\mathbf{x}_{i}, \mathcal{W}_{i}\right)\end {equation}

\begin{equation}\mathbf{x}_{i+1}=\mathcal{F}\left(\mathbf{x}_{i}, \mathcal{W}_{i}\right)+h\left(\mathbf{x}_{i}\right)\end{equation}
where $\mathcal{F}$ stands for the heterogeneous kernel-based residual block processing. Since  $h\left(\mathbf{x}_{l}\right)$ is an identity map, Eq 3 can be derived:

\begin{equation}\mathbf{x}_{i+1}=\mathcal{F}\left(\mathbf{x}_{i}, \mathcal{W}_{i}\right)+\mathbf{x}_{i}\end{equation}

\subsection{Gradient Cosine Similarity Loss Function}

To make the reconstructed image $I^{SR}$ obtained from the generator network closer to the high-resolution image $I^{HR}$, it is necessary to provide a neural network loss function with effective constraints. We chose the \hys{spatial gradient} of the image as the feature that measures the similarity between two images. When there is an edge in the image, there must be a high gradient value. Conversely, when there is a relatively smooth region in an image, the gray value changes little, and the corresponding gradient is also small. Using the gradient as a feature not only captures contours, images, and some texture information but also further weakens the effects of lighting. The gradient of an image at a pixel point $(x, y)$ is a vector with direction and size. $G_{x}$ is the gradient of $I$ in direction $X$, and $G_{y}$ is the gradient of $I$ in direction $Y$ direction. The gradient vector $\boldsymbol{v}$ can be expressed as Eq \ref{eq5}.

\begin{equation}
\boldsymbol{v}=\left[\boldsymbol{G}_{\boldsymbol{x}}, \boldsymbol{G}_{\boldsymbol{y}}\right]^{\boldsymbol{T}}
\label{eq5}
\end{equation}

The infrared images in the dataset are RGB images, which are three-channel images\cite{bradski2008learning}. The gradient between the high-resolution three-channel image $I^{HR}$ and the super-resolution reconstructed three-channel image $I^{SR}$ can be expressed as Eqs \ref{eq6} and \ref{eq7}.

\begin{equation}\mathbf{I}_{G}^{H R}=\left(\mathbf{I}_{G_{r}}^{H R}, \mathbf{I}_{G_{g}}^{H R}, \mathbf{I}_{G_{b}}^{H R}\right)
\label{eq6}
\end{equation}

\begin{equation}\mathbf{I}_{G}^{S R}=\left(\mathbf{I}_{G_{r}}^{H R}, \mathbf{I}_{G_{g}}^{H R}, \mathbf{I}_{G_{b}}^{H R}\right)
\label{eq7}
\end{equation}

$\mathbf{I}_{G}^{H R}$ indicates the gradient vector of the high-resolution image. The subscript of $G_{g}$ indicates \hys{the $green$ channel} of the high-resolution image. Other subscripts indicate \hys{different image channels of $red$ and $blue$.} For super-resolution reconstructed images $I^{SR}$, the subscript indicates the same. We use the cosine similarity to measure the similarity between these two vectors, as shown in Eq \ref{eq8}.

\begin{equation}\cos _{sim }(\mathbf{X}, \mathbf{Y})=\frac{\mathbf{X} \cdot \mathbf{Y}}{\|\mathbf{X}\| \cdot\|\mathbf{Y}\|}
\label{eq8}
\end{equation}

$\mathbf X$ and $\mathbf Y$ represent two matrices that can be multiplied by points. The high-resolution image gradient $\mathbf{I}_{G}^{H R}$ and the SR image gradient $\mathbf{I}_{G}^{S R}$ can be calculated according to algorithm 1.

We calculate the cosine similarity by stretching the two matrices into a one-dimensional vector. Likewise, the similarity between the high-resolution image gradient $\mathbf{I}_{G}^{H R}$ and the SR image gradient $\mathbf{I}_{G}^{S R}$ can be calculated according to algorithm 1. The generator loss function of the SRGAN and WGAN includes content loss and adversarial loss. The generator loss function of HetSRWGAN is shown in Eq \ref{eq10}:

\begin{algorithm}[t]{}
\caption{Gradient Cosine Similarity Loss Function} 
\hspace*{0.02in} {\bf Input:} 
$I^{SR}$ , $I^{HR}$\\
\hspace*{0.02in} {\bf Output:} 
Gradient Cosine Similarity
\begin{algorithmic}[1]
\State Infrared images can be processed into RGB images\cite{bradski2008learning}. 
\While{not convergent} 
\State $I^{HR}\longrightarrow \left(\mathbf{I}_{G_{r}}^{H R}, \mathbf{I}_{G_{g}}^{H R}, \mathbf{I}_{G_{b}}^{H R}\right)$
\State $I^{SR}\longrightarrow \left(\mathbf{I}_{G_{r}}^{S R}, \mathbf{I}_{G_{g}}^{S R}, \mathbf{I}_{G_{b}}^{S R}\right)$ \Comment {Gradient matrix.}
\State ${\bf{X}^{'}} = {\left[ {{\bf{I}}_{{G_r}}^{HR},{\bf{I}}_{{G_g}}^{HR},{\bf{I}}_{{G_b}}^{HR}} \right]_{1 \times m}}$ 
\State ${\bf{Y}^{'}} = {\left[ {{\bf{I}}_{{G_r}}^{SR},{\bf{I}}_{{G_g}}^{SR},{\bf{I}}_{{G_b}}^{SR}} \right]_{1 \times m}}$ \Comment{Matrix compression.}
\State $F_{\cos }\left(\bf{X}^{'}, \bf{Y}^{'}\right)=\frac{\bf{X}^{'} \cdot {{\bf{Y}^{'}}^T}}{\left\|\bf{X}^{'}\right\| \cdot\left\|\bf{Y}^{'}\right\|}$ \Comment{Cosine similarity.}
\EndWhile
\State \Return ${F_{\cos }}\left( {{\bf{X}^{'}},{\bf{Y}^{'}}} \right)$ 
\end{algorithmic}
\end{algorithm}

\begin{equation}\operatorname{Loss}^{S R}=l_{X}^{S R}+\lambda l_{G e n}^{S R}+\mu\left(1-F_{cos}\right)
\label{eq10}
\end{equation}

where $l_{X}^{S R}$ and $l_{Gen}^{S R}$ represent the content loss and adversarial loss, respectively.
\section{Experiments and Evaluations}\label{se.4}

\subsection{Training Details}

Following SRGAN, all experiments were performed with a scaling factor of (4, applied to the 2x2 image) between LR and HR images. We used the PSNR and structural similarity index (SSIM) to evaluate the reconstructed images. Super-resolved images were generated using the reference methods, including SRMD, IMDN, DPSR, DBPN, SRCNN, FSRCNN, ESPCN, SRGAN, and super-resolution Wasserstein GAN (SRWGAN). The generator was trained using the loss function presented in Eq \ref{eq10} with $\lambda=0.001$ and $\mu=0.001$. The learning rate was set to 0.0001. We observed that a larger batch size benefits training a deeper network. We set the batch size to 64. For optimization, we used Adam \cite{kingma_adam:_2014} with $\beta_{1}=0.9$ in the generator. For the WGAN, we used the Asynchronous SGD (ASGD)\cite{odena_faster_2016} in the discriminator. We implemented our models with the PyTorch framework and trained them using NVIDIA TITAN X (Pascal) GPUs.

\begin{figure}[t]
\centering
\includegraphics[width=0.8\textwidth]{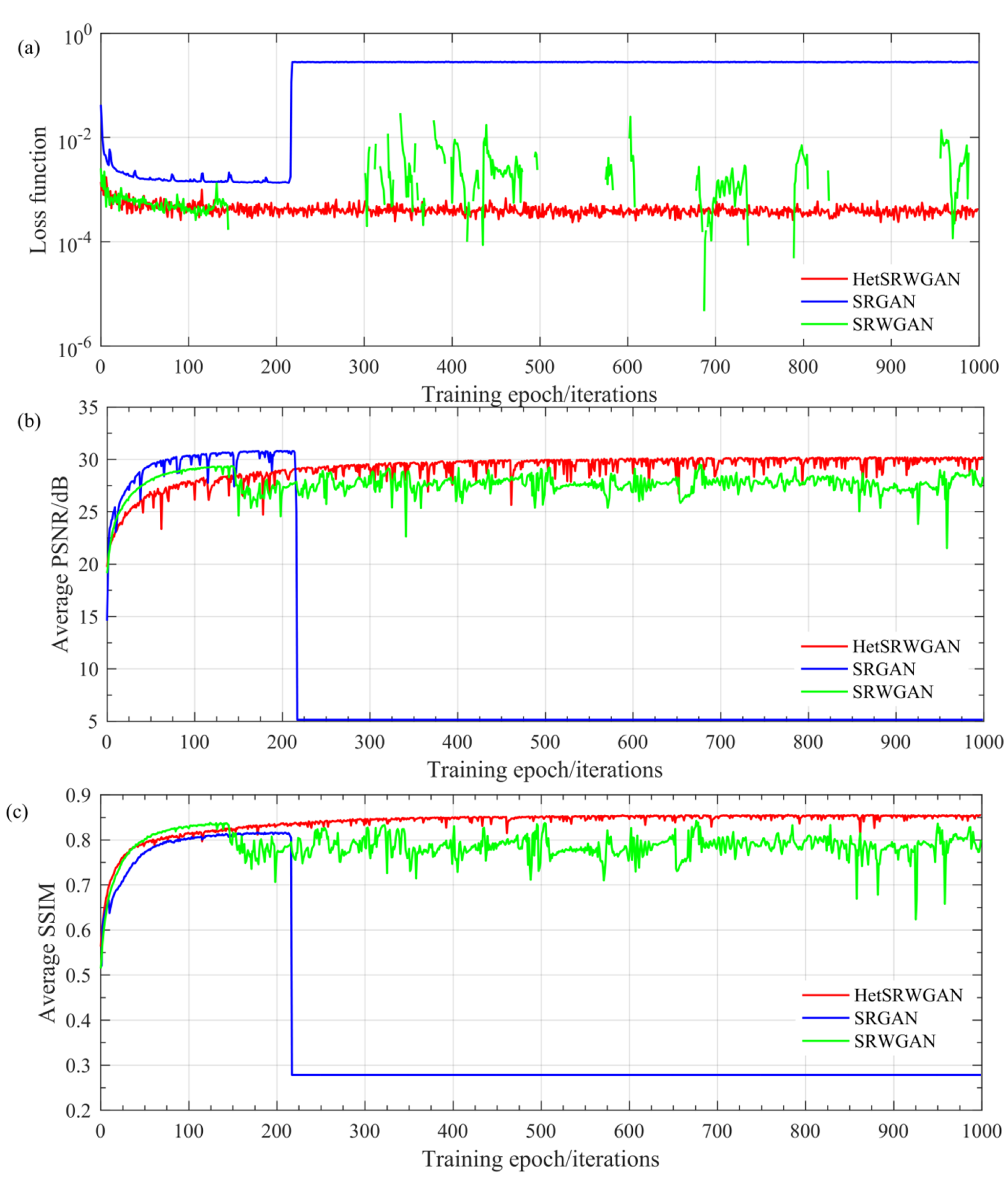}
\caption{(a): Changes of loss function with the number of iterations on the dataset CVC-09-1K (b): CVC-09-1K Dataset Training Average PSNR, (c): CVC-09-1K Dataset Training Average SSIM}
\label{Fig.3}
\end{figure} 

For training, we primarily used the CVC-09: FIR Sequence Pedestrian Dataset\cite{socarras2013adapting}. In CVC-09, a sequence is composed of two sets of images, the day and night sets, a designation which refers to the time of day at which they were acquired. The first set contains 5990 frames, the second set contains 5081 frames, and each sequence was divided into training and testing sets. We performed experiments on two datasets, namely, fusionA-22 and fusionC-22, which contain images obtained by fusing infrared and visible light, using the methods of literature\cite{liu_infrared_2017} and literature\cite{zhang_infrared_2017}, respectively\cite{huang_huang_2019}. \hys{An image after the fusion of IR and visible light images will have better visual quality, and it will be easier to distinguish details such as characters in the image. The fused image also maintains significant information from the infrared image but makes the performance of the algorithm more easily visualized.}
\begin{table}[t]
\centering
\caption{Quantitative evaluation of  SR algorithms: Average PSNR/SSIM for scale factors $\times 4$. SRGAN$^1$ has model collaps.}
\label{tab:my-table}
\begin{tabular}{@{}c|c|c|cccc@{}}
\toprule
\multirow{2}{*}{Algorithm} &
  \multirow{2}{*}{Params $\downarrow$} &
  \multirow{2}{*}{FLOPs $\downarrow$} &
  \multicolumn{2}{c}{PSNR/$dB$ $\uparrow$} &
  \multicolumn{2}{c}{SSIM $\uparrow$} \\ \cmidrule(l){4-7} 
 &
   &
   &
  \multicolumn{1}{c|}{fusionA-22} &
  \multicolumn{1}{c|}{fusionC-22} &
  \multicolumn{1}{c|}{fusionA-22} &
  fusionC-22 \\ \midrule
HetSRWGAN(Ours) & 0.496M & 0.095G & 30.302 & 31.987 & 0.858 & 0.883 \\
SRMD\cite{zhang2018learning}            & 1.552M & 0.063G & 33.210 & 33.850 & 0.834 & 0.852 \\
IMDN\cite{hui2019lightweight}            & 0.893M & 91.70G & 29.725 & 30.057 & 0.735 & 0.751 \\
DPSR\cite{zhang2019deep}            & 2.995M & 0.052G & 32.692 & 31.662 & 0.825 & 0.810 \\
DBPN\cite{haris2018deep}           & 10.41M & 0.106G & 17.438 & 17.934 & 0.816 & 0.842 \\
SRWGAN          & 0.956M & 0.132G & 28.319 & 28.520 & 0.799 & 0.805 \\
SRGAN$^1$\cite{ledig2017photo}       & 0.956M & 0.132G & 5.150  & 30.444 & 0.278 & 0.871 \\
SRCNN\cite{dong2015image}           & 0.148M & 0.182M & 29.437 & 30.170 & 0.754 & 0.789 \\
FSRCNN\cite{dong2016accelerating}          & 0.013M & 0.077M & 30.624 & 31.094 & 0.797 & 0.822 \\
ESPCN\cite{shi2016real}           & 0.061M & 0.001G & 30.814 & 31.607 & 0.789 & 0.819 \\ \bottomrule
\end{tabular}
\end{table}

\subsection{Performance of the Final Networks}

We compared the performance of three different super-resolution reconstruction algorithms based on generative adversarial networks. Since the GAN cannot simply use the loss function to judge the network training situation, we selected the image after the end of each batch of training to calculate the PSNR and SSIM values. When there are too many model parameters, mode collapse will occur. As the number of iterations increased, SRWGAN was more robust. SRGAN experiences mode collapse. Although the SRWGAN introduces gradient punishment to solve the problem that the network cannot be trained in the later stages, using cross-entropy as a loss function requires considerable time to adjust parameters and still cannot guarantee the stability of the model. Therefore, the loss function will have a negative value, which will cause the curve to be discontinuous. There was no situation where convergence or instability was not possible. The results are shown in Figure \ref{Fig.3}.

The total number of parameters for HetSRWGAN was reduced by 496657 compared to that for SRGAN, a reduction of 52\% (Table 1). The significantly reduced total number of parameters helps to reduce the computational complexity of the model and improve robustness.


\begin{figure}[t]
\centering
\includegraphics[width=0.6\textwidth ]{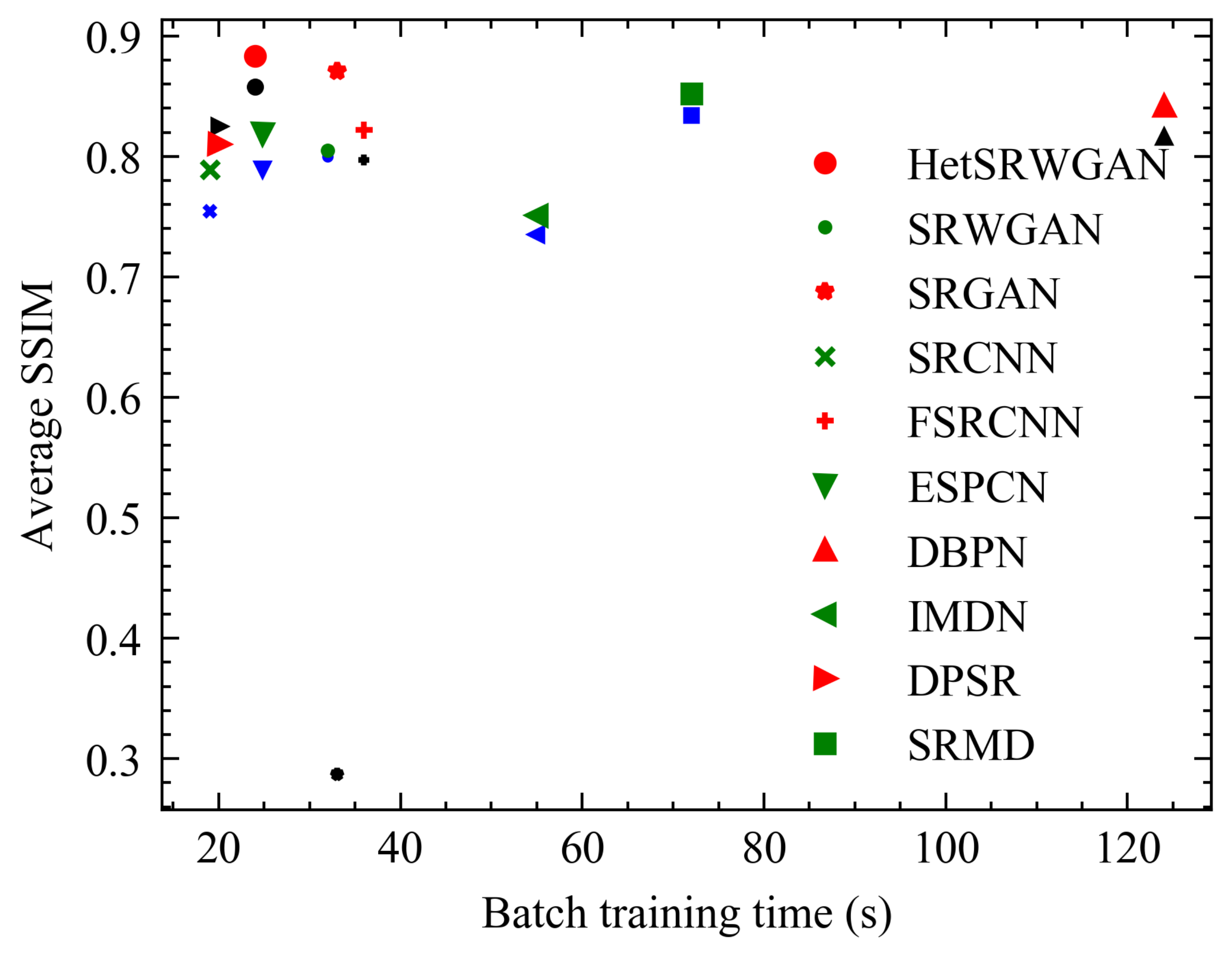}
\caption{Time efficiency comparison of all reconstruction methods. The same colour means the same method.The horizontal axis represents the time required for one training session, and the vertical axis represents an objective indicator after the model converges.}
\end{figure} 


\begin{figure}[t]
\centering
\includegraphics[width=0.8\textwidth]{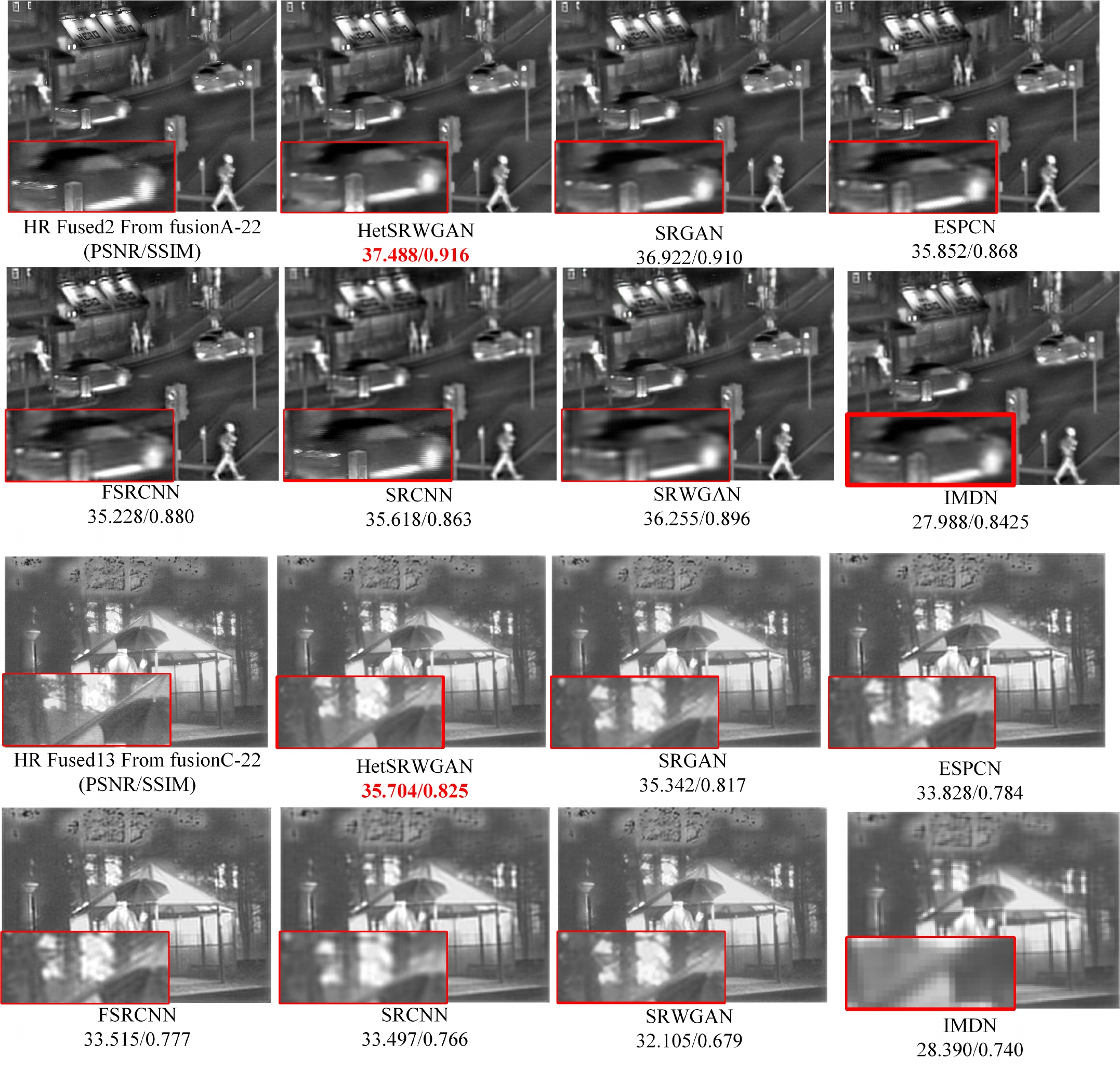}
\caption{Super-resolution image reconstruction effect comparison schematic diagram. From left to right: original HR image, HetSRWGAN, SRGAN, ESPCN, FSRCNN, SRCNN, SRWGAN, IMDN, Corresponding PSNR and SSIM are shown below the figure. \textcolor{red}{Red} indicates the best. [$\times 4$ upscaling] }
\end{figure}

The SRMD model obtains better performance based on the PSNR; however, it has a large number of parameters, resulting in long training and inference times and greater memory consumption. (Table 1, Figure 4) The objective evaluation indices of the average PSNR and average SSIM were calculated. DNNs have a good effect in reconstructing visible images, but because of the features of single-frame infrared images with few features and high redundancy, the reconstruction effect is not good. (Table 1)

SRGAN does not provide control of the generation process, and there is mode collapse (see Figure 1).
The new loss function and HetResidual block make the models faster to train and converge. The HetSRWGAN takes 24 seconds to train each batch, and the average SSIM is 0.858 and 0.883. (see Table 1) Compared with other methods, HetSRWGAN has the best time efficiency and average SSIM. Figure 5 shows the reconstructions produced by different algorithms. 

Figure 5 shows that our proposed HetSRWGAN outperformed previous approaches in both sharpness and amount of detail. Previous GAN-based methods sometimes introduce artifacts. For example, SRGAN adds noise to the entire image. HetSRWGAN removes these artifacts and produces natural results.

\section{Conclusions}\label{se.5}

Our proposed HetSRWGAN method can be well used for infrared image super-resolution reconstruction. We proposed a novel architecture composed of several heterogeneous kernel-based residual blocks without BN layers. A gradient cosine similarity loss function was developed, which can provide stronger supervision of image details, such as edges, and the reconstructed high-resolution images contain more details and realistic textures.

\section*{Acknowledgement}
This research supported by the Nature Science Foundation of China grants No.61876049, and No.61762066.
%
%

\bibliographystyle{splncs04.bst}
\end{document}